\documentclass[prb,twocolumn,superscriptaddress,showpacs,amsmath,amssymb]{revtex4}

\usepackage{graphicx}
\usepackage{latexsym}
\usepackage{amsmath}
\usepackage{amssymb}
\usepackage{amsfonts}
\usepackage{color}
\usepackage{bm}
\usepackage{bbm}

\usepackage{verbatim}

\begin{document}

\title{Comment on the paper by D. Efremov and Yu.N. Ovchinnikov "Singular ground state of multiband inhomogeneous superconductors", Phys. Rev. B {\bf 99}, 224508 (2019)}

\author{Mihail Silaev}
  \affiliation{Department of
Physics and Nanoscience Center, University of Jyv\"askyl\"a, P.O.
Box 35 (YFL), FI-40014 University of Jyv\"askyl\"a, Finland}

\author{Thomas Winyard}
\affiliation{School of Mathematics, University of Leeds,Leeds LS2 9JT, England}

\author{Egor Babaev}
\affiliation{Department of Theoretical Physics, The Royal
Institute of Technology, Stockholm SE-10691, Sweden}

 \begin{abstract}
We show that the conclusion reported in Ref. \cite{ovchinnikov2019singular}, that there are no spontaneous magnetic fields in multiband superconductors that break time reversal symmetry, is incorrect. We demonstrate that the state proposed in Ref. \cite{ovchinnikov2019singular} is not a solution of the Ginzburg-Landau equations for the considered model. The reason is that in Ref. \cite{ovchinnikov2019singular}
one of the Ginzburg-Landau equations is neglected and substituted by the spurious zero current restriction. This restriction together with all of the Ginzburg-Landau equations leads to an overdetermined system which does not have a solution. This inconsistency invalidates all the results of the paper \cite{ovchinnikov2019singular}.  
 \end{abstract}

 \maketitle
\section{Introduction}
Recent experiments have reported evidence for $s+is$ or $s+id$ superconductivity
in K$_x$Fe$_2$As$_2$ \cite{Grinenko2017,grinenko2018emerging}.
This type of superconductivity should arise in spin-singlet multiband superconductors where the broken time reversal symmetry (BTRS) is due to interband interactions. A number of theoretical arguments have been advanced in favour of these states\cite{Lee.Zhang.Wu:09,Maiti2013,Johan,StanevTesanovic,Boeker2017}. The evidence was based on the observation of a superconducting phase which has spontaneous magnetic field below a certain critical temperature. The emergence of such spontaneous magnetic field has been discussed for 
$s+id$  \cite{Lee.Zhang.Wu:09,ChubukovMaitiSigrist,lin2016distinguishing,
Silaev.Garaud.ea:15,2016PhRvL.116i7002G,vadimov2018polarization,domainwalls} and $s+is$
 \cite{Garaud2014,Silaev.Garaud.ea:15,2016PhRvL.116i7002G,lin2016distinguishing,vadimov2018polarization,garaud2018properties,domainwalls}  superconductors.
 
Two sources of magnetic field have been considered: inhomogeneous external potentials such as impurities or temperature gradients, and domain walls.
In contrast to the well studied $p+ip$ case, spontaneous magnetic
fields in $s+is$ or $s+id$ systems have received less attention and even some inconsistent claims in the literature, even from papers arguing in favour of spontaneous magnetic field. For example, a paper has recently claimed that spontaneous magnetic field appears only in the $s+id$ state and not in the $s+is$ state \cite{ChubukovMaitiSigrist}, while another paper claimed that spontaneous magnetic fields appear in both states but that $s+id$ exhibits a much stronger response\cite{lin2016distinguishing}. A number of counterexamples
exist for both of these statements\cite{domainwalls,Garaud2014,garaud2018properties,vadimov2018polarization}, for an example that specifically counters both see \cite{domainwalls}. It was demonstrated that spontaneous magnetic field can be generated from two properties (i) in an isotropic system non-collinear gradients of the relative densities and relative phase between components generate magnetic field (see detailed discussion in \cite{Garaud2014,garaud2018properties}
 and (ii) in an anisotropic system, the different anisotropies for different 
 components rather generically mix the magnetic mode with other modes \cite{silaev2017non,winyard2018,vadimov2018polarization,speight2019chiral,winyard2018skyrmion,Silaev.Garaud.ea:15,2016PhRvL.116i7002G}. The consequence of the ideas explored above is that a weakly inhomogeneous $s+is$ and $s+id$ system will spontaneously generate magnetic fields.
   
In a recent paper by D. Efremov and Yu. Ovchinnikov \cite{ovchinnikov2019singular} the claim was advanced that multiband BTRS superconductors do not have spontaneous magnetic field. In this response we will demonstrate that this assertion is incorrect and demonstrate the error in their method and assumptions.

If the reader wants to see work that demonstrates the existence of spontaneous magnetic field in BTRS systems see \cite{domainwalls, speight2019chiral, vadimov2018polarization}.

\section{Model}
\subsection{Free energy}
We will mostly use the notation used in the paper \cite{ovchinnikov2019singular}, however where there are slight deviations that we will make clear. We consider the example of an $s+id$ superconductor, however our arguments hold for $s+is$ and $p+ip$ systems. It is described by the two-component GL free energy  
 \begin{align}
 & F = \int d^3r  [f_0 + f_c + B^2/8\pi] \label{Eq:F}
 \\ 
 & f_0 = \sum_{i=1}^2 
 ( \alpha_i|\Psi_i|^2 + 
 \frac{\beta_i}{2} |\Psi_i|^4 + 
 \frac{\hbar^2}{2m_i} 
 |\partial_- \Psi_i|^2 )
 \\
 & f_c= \frac{\gamma_3}{4} 
 |\Psi_1\Psi_2|^2 + 
 \frac{\gamma_2}{2}  (\Psi_1^*\Psi_2)^2 \nonumber\\
 & + \frac{\hbar^2}{4 m_c}
   [(\partial_- \Psi_1)^*_x(\partial_- \Psi_2)_x - 
  (\partial_- \Psi_1)^*_y(\partial_- \Psi_2)_y  ]
  + c.c.
 \end{align}
Where $\partial_- = \nabla - 2ie \bm A/\hbar c$ is the covariant derivative with respect to the $U(1)$ gauge field $\boldsymbol{A}$. We note that whenever using indices, we will use Greek indices for spatial coordinates and Latin indices for component coordinates. Additionally, repeated Greek (spatial) indices  are summed over. Note that this is the opposite from the authors usual convention however matches the convention used in \cite{ovchinnikov2019singular}. $\Psi_i$ are complex order parameters which, following EO, we write $\Psi_k = |\Psi_k| e^{i\phi_k}$. This leads to the GL equations
 \begin{align} \label{Eq:GL}
 &\frac{\delta F }{ \delta |\Psi_k| }  =0, \quad \quad 
  \boldsymbol{\nabla}\cdot \boldsymbol{j} = 0
 \\ \label{Eq:GLPhase}
 &\frac{\delta F }{ \delta \varphi_{12} } =0, \quad \quad
 \end{align}
 where $\varphi_{12} = \phi_1-\phi_2$ is the phase difference between two order parameter components and $\bm j$ is the current. Note that for any configuration to be a minimiser of Eq. (\ref{Eq:F}), either local or global, it must be a critical point of {\emph all} of the above GL equations.
 
Efremov and Ovchinnikov (EO) considered $s+id$ superconductors described by the functional (\ref{Eq:F})  with spatially-dependent coefficients 
$\alpha_i =  \alpha_{i}^{(0)} + \delta \alpha_i (\bm r)$, where $\delta \alpha_i$ is small and can be considered as a perturbation. The authors claim to find a state (that later we call the EO state) which has zero current density $\bm j=0$ and thus zero magnetic field response ${\bm B} = 0$, which they also claim is the ground state of the free energy functional in Eq. (\ref{Eq:F}). 

To find EO state the authors of Ref.\cite{ovchinnikov2019singular} took into account 
only Eqs.(\ref{Eq:GL}) supplemented by the spurious zero-current restriction $\bm j=0$. They neglected Eq.(\ref{Eq:GLPhase}) which is not satisfied by the EO state as shown below. Therefore this state is not a ground nor metastable state of the free energy functional.

In fact we will show that no such state exists that satisfies all the GL equations and the spurious zero-current restriction ${\bm j} = 0$.
 
We now introduce some useful nomenclature $\bm p_k = \nabla\phi_k - 2e \bm A/\hbar c$ for simplicity, which leads to $\partial_- \Psi_k = 
(\nabla |\Psi_k| + i \bm p_k |\Psi_k| )e^{i\varphi_k}$. The free energy can now be written,
  \begin{align}
  & f_0 = \sum_{k=1}^2 
 ( \alpha_k|\Psi_k|^2 + 
 \frac{\beta_k}{2} |\Psi_k|^4 + 
 \frac{\hbar^2}{2m_k}  ( |\nabla \Psi_k|^2  + p_k^2 |\Psi_k|^2 )
 \\
 &\nonumber f_c= 
 \frac{1}{2}|\Psi_1\Psi_2|^2
 [ \gamma_3 + 2\gamma_2 \cos(2\varphi_{12}) ]  
  \\
 &\nonumber +\frac{\hbar^2}{2 m_c}\cos\varphi_{12}
   K_{\alpha} \left[
    \nabla_\alpha |\Psi_1| \nabla_\alpha |\Psi_2| +
    |\Psi_1| |\Psi_2|  p_{1\alpha}p_{2\alpha}
      \right]
   \\
  & +\frac{\hbar^2}{2 m_c}\sin\varphi_{12}
      K_{\alpha} [
      p_{2\alpha} |\Psi_2|\nabla_\alpha |\Psi_1| - 
      p_{1\alpha}|\Psi_1| \nabla_\alpha |\Psi_2| 
      ] 
       \end{align} 
  where $K_{x}=-K_{y} =1$.

\subsection{GL equations}

 In our chosen variables  the current density is given by 
 \begin{align}\label{Eq:j}
 &\frac{\hbar c}{2e } j_\alpha = \sum_{k=1}^2 \frac{\hbar^2}{m_k}  
   p_{k\alpha} |\Psi_k|^2 
  + 
  \\ \nonumber
  & \frac{\hbar^2}{2m_c} \cos\varphi_{12} |\Psi_1||\Psi_2| 
  \delta_{\alpha\beta}K_{\beta} (p_{1\beta} + p_{2\beta}
  )+
  \\ \nonumber
  & \frac{\hbar^2}{2m_c} \sin\varphi_{12} \delta_{\alpha\beta}K_{\beta}
   (\nabla_\beta |\Psi_1||\Psi_2| 
   -\nabla_\beta |\Psi_2||\Psi_1|) 
   \end{align}

  
The GL equations $\delta F/\delta |\Psi_{1,2}|=0$ can be  written as
\begin{widetext}
  \begin{align} \label{Eq:GLPsi1}
  & 2 \alpha_1 |\Psi_1| + 2\beta_1 |\Psi_1|^3 + 
  [\gamma_3 +2\gamma_2\cos(2\varphi_{12})] |\Psi_1| |\Psi_2|^2 
  +
 \frac{\hbar^2}{m_1} (p_1^2|\Psi_1| - \nabla^2 |\Psi_1|)
 + 
 \\ \nonumber
 & \frac{\hbar^2}{2m_c} K_{\alpha}
 [\cos\varphi_{12}\, p_{1\alpha}p_{2\alpha}|\Psi_2| - 
 \nabla_\alpha (\cos\varphi_{12} \nabla_\alpha |\Psi_2|) ] 
 - 
 \frac{\hbar^2}{2m_c} K_{\alpha} [\sin\varphi_{12} p_{1\alpha}\nabla_\alpha |\Psi_2|  + \nabla_\alpha(\sin\varphi_{12} p_{2\alpha} |\Psi_2| )] =0
 \\   \label{Eq:GLPsi2}
  & 2 \alpha_2 |\Psi_2| + 2\beta_2 |\Psi_2|^3 + 
  [\gamma_3 +2\gamma_2\cos(2\varphi_{12})] |\Psi_2| |\Psi_1|^2 
  +
 \frac{\hbar^2}{m_2} (p_2^2|\Psi_2| - \nabla^2 |\Psi_2|)
 + 
 \\ \nonumber
 & \frac{\hbar^2}{2m_c} K_{\alpha}
 [\cos\varphi_{12}\, p_{1\alpha}p_{2\alpha}|\Psi_1| - 
 \nabla_\alpha (\cos\varphi_{12} \nabla_\alpha |\Psi_1|) ] 
 +
 \frac{\hbar^2}{2m_c} K_{\alpha} 
 [\sin\varphi_{12} p_{2\alpha}\nabla_\alpha |\Psi_1|  + 
 \nabla_\alpha(\sin\varphi_{12} p_{1\alpha} |\Psi_1| )] =0
  \end{align}
 The GL Eq.(\ref{Eq:GLPhase}) corresponding to the condition $\delta F/\delta \varphi_{12}=0$ reads
  \begin{align} \label{Eq:GLPhi12}
  & -\frac{2m_c}{m_1}\nabla_\alpha (p_{1\alpha}|\Psi_1|^2) + \frac{2m_c}{m_2}\nabla_\alpha (p_{2\alpha}|\Psi_2|^2) - \frac{2m_c}{\hbar^2} \gamma_2 |\Psi_1 \Psi_2|^2\sin(2\varphi_{12}) - 
   \sin\varphi_{12} K_{\alpha}
   (\nabla_\alpha|\Psi_1|\nabla_\alpha|\Psi_2| + p_{1\alpha}p_{2\alpha} |\Psi_1\Psi_2| ) +
   \\ \nonumber
  & \cos\varphi_{12}K_{\alpha}
   (p_{2\alpha} |\Psi_2| \nabla_\alpha |\Psi_1| - p_{1\alpha} |\Psi_1| \nabla_\alpha |\Psi_2| ) +
  K_{\alpha} \nabla_\alpha (\sin\varphi_{12}  
  (|\Psi_1|  \nabla_\alpha |\Psi_2| + 
  |\Psi_2|  \nabla_\alpha |\Psi_1| )) + 
  \\ \nonumber
  & K_{\alpha} \nabla_\alpha [\cos\varphi_{12}  (p_{2\alpha} - p_{1\alpha}) |\Psi_1\Psi_2|] =0.
  \end{align}
  \end{widetext}
 The equation (\ref{Eq:GLPhi12}) is neglected in Ref.\cite{ovchinnikov2019singular}.
 Below we show that the full system 
 (\ref{Eq:GLPsi1},\ref{Eq:GLPsi2},\ref{Eq:GLPhi12}) is inconsistent with the restriction of $\bm j=0$ where the current is given by Eq.(\ref{Eq:j}).
  
 
\section{Linearisation and EO state inconsistency proof}

We follow EO and consider a weak inhomogeneity or the tail of an inhomogeneity
$\alpha_i =  \alpha_{i}^{(0)} + \delta \alpha_i (\bm r)$, where $|\delta \alpha_i|\ll |\alpha_{i}^{(0)}|$. This allows considering perturbations about the ground state where the linear contribution is the most important.

We linearise around the ground state values $|\Psi_k| = u_k$, $\varphi_{12} = \frac{\pi}{2}$ and $\boldsymbol{p}_k = 0$, writing our fields as $\left|\Psi_k\right| = u_k + \tilde{\Psi}_k$ and $\varphi_{12} = \frac{\pi}{2} + \tilde{\varphi}_{12}$. We then expand in these fields assuming that $\tilde{\Psi}_k$, $\tilde\varphi_{12}$ and $\boldsymbol{p}_k$ are small and neglecting any quadratic or higher order terms.

We first linearise the zero current condition $\boldsymbol{j} = 0$ where $\boldsymbol{j}$ is given in Eq. (\ref{Eq:j}). This can be split into two conditions by noting that the equation is polarised, hence considering $\boldsymbol{r}\cdot\boldsymbol{j} = 0$ and $\left.\boldsymbol{r}\times\boldsymbol{j}\right|z = 0$ we find the following two linearised conditions,  
  \begin{align} \label{Eq:Restr1}
  & u_2\nabla \tilde{\Psi}_1 = u_1\nabla \tilde{\Psi}_2
  \\ \label{Eq:Restr2}
  & \sum_{k=1}^2 \frac{1}{m_k}  
  \bm p_{k} u_k^2 =0.
  \end{align}  
These restrictions coincide with Eqs.(20,21) from Ref.\cite{ovchinnikov2019singular}. Also note that as we assume, at infinite distance from the defect, the fields decay to their ground state values, Eq. (\ref{Eq:Restr1}) can be rewritten as $u_2 \tilde{\Psi}_1 = u_1\tilde{\Psi}_2$. 
 
Using the restrictions (\ref{Eq:Restr1},\ref{Eq:Restr2})
 we can eliminate the variable $\tilde\Psi_2$ and express $\bm p_2$
 and $\bm p_1$ in terms of the phase difference
\begin{align}
 \boldsymbol{\nabla} \tilde{\varphi}_{12} = \left(1 - \frac{u_1^2}{u_2^2}\frac{m_2}{m_1}\right)\boldsymbol{p}_1
 \end{align}

  Then Eqs.(\ref{Eq:GLPsi1},\ref{Eq:GLPsi2}) are reduced to
 %
{
\begin{align} \label{Eq:Lin1}
  & \frac{m_1}{\hbar^{2} m_2} ( 2\tilde\alpha_1 u_1
  + a_1 \tilde\Psi_1 ) = 
  \frac{\nabla^2 \tilde\Psi_1}{m_2} -
  \frac{u_1^2u_2 m_1}{u_2^2m_1-u_1^2m_2} \frac{K_{\alpha} \nabla^2_\alpha \tilde\varphi_{12}}{2m_c}
   \\ \label{Eq:Lin2}
   & \frac{1}{\hbar^{2}}( 2\tilde\alpha_2 u_1
  + a_2 \tilde\Psi_1 ) = 
  \frac{\nabla^2 \tilde\Psi_1}{m_2} -
  \frac{u_1^2u_2 m_1}{u_2^2m_1-u_1^2m_2} \frac{K_{\alpha} \nabla^2_\alpha \tilde\varphi_{12}}{2m_c}
\end{align} 
}
 where we denote $a_1 = 2\alpha^{(0)}_1 + 6 \beta_1 u_1^2 + (\gamma_3 - 2\gamma_2) (u_1^2 + 2u_2^2 )$ 
 and 
 $a_2 = 2\alpha^{(0)}_2 + 6 \beta_2 u_2^2 + (\gamma_3 - 2\gamma_2) (u_2^2 + 2u_1^2 )$.
 At the same time Eq.(\ref{Eq:GLPhi12}) yields
 
{
\begin{align} \label{Eq:Phi12Lin}
 2u_2 K_{\alpha} \nabla_\alpha^2 \tilde \Psi_1 = 
 \frac{4 m_c u_1^2u_2^2}{m_1u_2^2-m_2u_2^2} \nabla^2 \tilde\varphi_{12} -
  \frac{4 m_c}{\hbar}\gamma_2 u_1^2 u_2^2 \tilde{\varphi}_{12}
 \end{align}} 
 As we have said before, the EO assumption has given an overdetermined system of equations, this is more apparent now as we have three independent equations 
 (\ref{Eq:Lin1},\ref{Eq:Lin2},\ref{Eq:Phi12Lin}) for two fields $\Psi_1$ and
 $\tilde\varphi_{12}$, which will clearly be inconsistent in general. 
 
Let us show this explicitly for the particular case when the impurity is axially symmetric, that is $\tilde \alpha_k (\bm r) = \tilde \alpha_k (r_\perp)$ where $r_\perp= \sqrt{x^2+y^2}$. 
 From Eqs. (\ref{Eq:Lin1},\ref{Eq:Lin2})
 we get that 
 \begin{equation}
 \tilde \Psi_1 = 2 
  \frac{(\tilde \alpha_1 m_1 - m_2 \tilde\alpha_2 )
  } {m_2a_2-m_1a_1}  u_1\label{Eq:Phi1}
  \end{equation}
From this it follows that $\tilde \Psi_1= \tilde \Psi_1 (r_\perp)$ so that $K_\alpha \nabla_\alpha^2\tilde \Psi_1 = 
\cos(2\theta) (\nabla_r^2 \tilde \Psi_1 - \frac{1}{r} \nabla_r \tilde \Psi_1)$ , where $\theta$ is the spatial polar angle.

Thus the solution of (\ref{Eq:Phi12Lin}) has the form $\tilde\varphi_{12} = \cos(2\theta)\tilde\varphi (r_\perp) $. If we take this ansatz and place it into Eqs.(\ref{Eq:Lin1},\ref{Eq:Lin2}) we will obtain a term proportional to
\begin{align}
&\nonumber K_\alpha \nabla_\alpha^2 [\cos(2\theta)\tilde\varphi (r_\perp)] = \cos^2(2\theta)(\tilde\varphi'' - \frac{1}{r}\varphi') + 4\sin^2(2\theta)\tilde\varphi' \\&+ \frac{1}{r}\left(3\cos(4x) + 1\right) \tilde\varphi.
\end{align}
As the above has both a $\cos (2\theta)$ and $\cos (4\theta)$ component, the result must be non-radial. The rest of the equation, namely terms proportional to $\nabla^2 \tilde \Psi_1$, $\tilde \Psi_1$ and $\alpha_i$, are all radial, hence this equation is general has only the solution $\tilde\varphi = 0$. This means that the phase difference doesn't fluctuate. 

Hence finally we can use Eqs. (\ref{Eq:Lin1}, \ref{Eq:Phi12Lin}) to get,
\begin{align}
&\nabla_r^2 \tilde \Psi_1 - \frac{1}{r} \nabla_r \tilde \Psi_1 = 0\\
&\nabla_r^2 \tilde \Psi_1 + \frac{1}{r} \nabla_r \tilde \Psi_1 = \frac{m_1}{\hbar^{2}} ( 2\tilde\alpha_1 u_1
  + a_1 \tilde\Psi_1 )
\end{align}
Which along with the condition for $\tilde\Psi_1$ in Eq. (\ref{Eq:Phi1}) and the requirement that far from the defect $\tilde\alpha_i \rightarrow 0$ leads to the relations
\begin{equation}
\tilde \alpha_1 = \frac{a_1}{a_2}\tilde\alpha_2 = \frac{m_2}{m_1}\tilde\alpha_2
\end{equation}
If we substitute the above into Eq. (\ref{Eq:Lin1},\ref{Eq:Phi1}) it leads to $\tilde\Psi_1 = 0$ and hence none of the fields fluctuate and that $\tilde\alpha_k = 0$ and there can be no defect. This is a contradiction as we have assumed there is a defect as EO did in Ref. \cite{ovchinnikov2019singular}. Hence there is no solution and the EO state cannot in general satisfy one of the Ginzburg Landau equations, namely Eq. \ref{Eq:GLPhase}.

Additionally we have shown that in general there is no solution that can satisfy all the Ginzburg Landau equations as well as the spurious zero current restriction and EOs conclusion that the ground state around a defect has no spontaneous magnetic field is incorrect.

Finally in the case where there is no defect, we have shown that in general there can be no fluctuations in any of the fields for there to be zero magnetic field. This means that in general there must be a magnetic response from defects, domain walls or fluctuations in any of the fields. 

\section{Conclusions}
We have  shown that the proposed EO state in Ref. \cite{ovchinnikov2019singular} is not in general a solution of  the Ginzburg Landau equations. This is due to the assumption $\boldsymbol{B} = 0$ and $\boldsymbol{j} = 0$ leading to an overdetermined system, that when followed through (as shown above) leads to an inconsistency in general. We showed this explicitly for a radial defect as an example, where we demonstrated that no EO state exists in the presence of a defect.

It is important to note that it may be possible to carefully construct a defect such that the magnetic response is zero in such a system, but this will be a fine tuned zero measure case.

We have also shown that fluctuations in any of the fields (even with no defect) will lead to spontaneous magnetic field.

It is also important to note that while we have focussed here on an $s+id$ system (as it was the chosen system for EO) our argument holds for both $s+is$ and $p+ip$ states also. The introduction of anisotropies leads to the coupling of fields in general as discussed in Ref. \cite{silaev2017non,winyard2018,speight2019chiral,domainwalls}.

\section{Acknowledgements}
We thank Martin Speight for useful discussions.
The work of TW is supported by the
UK Engineering and Physical Sciences Research Council
through grant EP/P024688/1. EB is supported by
the Swedish Research Council Grants No. 642-2013-7837,
VR2016-06122 and G{\"o}ran Gustafsson Foundation for Research
in Natural Sciences and Medicine.

\bibliography{bibliography}

 \end{document}